\documentclass{article}

\usepackage{arxiv}

\usepackage[utf8]{inputenc}
\usepackage[T1]{fontenc}
\usepackage{lmodern}
\usepackage{hyperref}
\usepackage{url}
\usepackage{booktabs}
\usepackage{amsfonts}
\usepackage{amssymb}
\usepackage{amsmath}
\usepackage{amsthm}
\usepackage{nicefrac}
\usepackage{microtype}
\usepackage{graphicx}
\usepackage[numbers]{natbib}
\usepackage{doi}
\usepackage{mathtools}
\usepackage{bm}

\title{A Detector-Based Inference Framework for Quantum Theory and Spacetime Geometry}

\author{Marcello Rotondo}

\renewcommand{\shorttitle}{Inference Framework for Spacetime Geometry}

\hypersetup{
pdftitle={A Detector-Based Inference Framework for Quantum Theory and Spacetime Geometry},
pdfauthor={Marcello Rotondo},
pdfsubject={quantum foundations, information geometry, spacetime geometry, quantum inference, emergent gravity},
pdfkeywords={quantum foundations, information geometry, spacetime geometry, quantum inference, emergent gravity}
}

\begin{document}
\maketitle
\begin{center}
Independent researcher

\texttt{marcello@gravity.phys.nagoya-u.ac.jp}
\end{center}

\begin{abstract}
We develop a detector-based framework in which quantum theory and spacetime geometry arise within a common inferential structure. Detector states and a detector kernel assign amplitudes to measurement events, allowing quantum theory to be interpreted as weighting hypothetical configurations consistent with observed detector clicks.

Using a Gaussian detector model with phase structure, we show that distinguishability induces an information geometry on detector-state space, described by the quantum geometric tensor. A Lorentzian spacetime metric is reconstructed from coupled position and time detector sectors, with both amplitude and phase deformations contributing to geometry. Scalar curvature acquires an operational interpretation as a local deficit of distinguishable outcomes.

We construct an effective consistency functional combining detector-deformation cost with a geometric term selected by locality and diffeomorphism invariance. Its stationary configurations yield the Einstein equation, with a stress-energy tensor arising from detector deformations. Vacuum configurations need not be flat, while local deformations provide an operational notion of matter and recover standard field-theoretic behavior in the scalar sector.
\end{abstract}

\keywords{
quantum foundations \and information geometry \and spacetime geometry \and quantum inference \and emergent gravity
}

\section{Introduction}

``It is wrong to think that the task of physics is to find out how nature is. Physics concerns what we can say about nature.'' This remark by Niels Bohr \cite{Bohr1958} captures a central feature of quantum theory: its formalism is concerned with the organization of statements about measurement outcomes, rather than with a direct description of underlying reality. In contrast, general relativity is typically formulated as a theory of spacetime itself, in which the metric tensor $g_{\mu\nu}$ is taken to represent an objective geometrical structure. Reconciling these two perspectives---operational in quantum theory and geometric in gravitation---remains one of the central conceptual challenges in theoretical physics.

A number of approaches have emphasized the role of information in physical theory. Bayesian and operational interpretations of quantum mechanics describe the quantum state as a tool for assigning probabilities to measurement outcomes rather than as a physical entity \cite{FuchsSchack2013}. Information geometry provides a mathematical framework for quantifying distinguishability between statistical models, both in classical and quantum settings \cite{AmariNagaoka2000,BraunsteinCaves1994,Petz1996}. In a different direction, Jacobson's derivation of the Einstein equation from thermodynamic considerations suggests that gravitational dynamics may be understood as an effective consistency condition rather than as a fundamental microscopic law \cite{Jacobson1995}. Verlinde's entropic interpretation of gravity pushes this line further, though in a framework conceptually distinct from the one pursued here \cite{Verlinde2011}. Relational and observer-dependent approaches, including those associated with Rovelli, similarly emphasize that physical structure should be defined through operationally meaningful relations rather than by appeal to an absolute background \cite{Rovelli1996}. At a more formal level, several developments in high-energy theory have emphasized that geometrical structures may be reconstructed from more primitive informational or kinematical data. Notable examples include the emergence of spacetime geometry from quantum entanglement in holographic duality \cite{Maldacena1997,RyuTakayanagi2006,VanRaamsdonk2010}, as well as derivations of gravitational dynamics from entanglement or thermodynamic considerations \cite{Jacobson2015}. While these approaches differ substantially in both setting and interpretation, they share with the present work the idea that spacetime geometry need not be taken as fundamental.

Despite these developments, a unified framework in which both quantum theory and spacetime geometry are grounded directly in detector responses is still lacking. In particular, while quantum theory is often interpreted operationally, the geometrical structures of general relativity are rarely tied in a comparably explicit way to the structure of measurement. The present work explores the possibility that such a connection can be established by taking detector outcomes as the fundamental starting point.

The guiding idea is to formulate physical theory in terms of the response structure defined by detectors. Rather than assuming a pre-existing spacetime in which events occur, we consider a space of possible detector outcomes, each corresponding to a possible measurement event. The primary object is then a detector response kernel relating hypothetical configurations to observable outcomes. Within this setting, the aim is to investigate how familiar structures of quantum theory and spacetime geometry can arise from the organization of these responses.

A central role is played by distinguishability. If physical statements are ultimately about detector outcomes, then the ability to distinguish neighboring outcomes is the natural candidate from which geometric structure should be built. This suggests that a spacetime metric might be reconstructed from a suitable notion of detector-state distinguishability. Once this is done, the next question is whether curvature and dynamics can be interpreted as consistency conditions governing the patching of local detector calibrations.

The purpose of this paper is to develop a framework along these lines. We first introduce a detector-based formulation in which amplitudes are associated with detector clicks rather than with the properties of underlying systems. We then examine how the path-integral structure of quantum theory can be reinterpreted as inference over hypothetical histories compatible with observed outcomes. Next, we construct the information geometry of detector states and show how a Lorentzian spacetime metric can be reconstructed from coupled spatial and temporal detector sectors. We will start from the vacuum sector, in which detector families remain in a reference Gaussian class, and then move to the non-vacuum sector, in which local deformations of the detector family modify the induced geometry. Finally, we introduce a global consistency functional governing admissible detector configurations and derive the associated effective field equations.

Two points should be emphasized from the outset. First, the framework is explicitly effective. It does not claim a microscopic derivation of spacetime from an underlying theory of detectors. Rather, it identifies the structures that arise when detector responses admit a smooth geometric description. Second, the gravitational sector is not claimed to be derived microscopically from the detector kernel alone. Instead, once a Lorentzian metric has been reconstructed from detector distinguishability, the Einstein--Hilbert term is argued to be the unique leading local scalar consistency term in the low-energy, diffeomorphism-invariant effective description. In this sense, the framework does not derive general relativity from first principles, but it clarifies why the Einstein–Hilbert action appears as the universal leading term once detector distinguishability is taken as primary.

The resulting picture is neither purely instrumentalist nor naively realist, but closer to a structural or relational viewpoint in which physical content is encoded in the relations between measurement outcomes and detector responses \cite{Rovelli1996,Ladyman1998}. The microscopic ontology is left open. What is taken as primary is the detector-response structure. The spacetime metric is then interpreted realistically, but only as an effective structure reconstructed from detector distinguishability. This places the framework closer to a form of structural realism than to either straightforward instrumentalism or a fundamental ontology of spacetime points.

\section{Detector Framework, Gaussian Reference Family, and Metric Reconstruction}

The central premise of the present framework is that the empirical content of physical theory is exhausted by detector outcomes. Rather than treating spacetime as a pre-existing arena in which events occur, we adopt the operational viewpoint that spacetime events are identified with detector clicks. A point $x$ in spacetime is therefore not an ontological location, but a label for a possible measurement outcome.

This perspective leads naturally to a reformulation of quantum theory in terms of detector response functions. Instead of assigning amplitudes to the presence of a particle at a given position, we assign amplitudes to the event that a detector labeled by $x$ registers a click. In this sense, the wavefunction is interpreted as encoding the response of a detector rather than the properties of an underlying system.

We formalize this idea by introducing a detector kernel
\begin{equation}
K(x,y),
\end{equation}
which assigns a response amplitude for a detector labeled by $x$ to click, conditioned on a parameter value $y$. The variable $y$ should be understood as parametrizing hypothetical configurations entering the inference problem, rather than as directly observable quantities. The associated probability density is
\begin{equation}
P(x \mid y) = |K(x,y)|^2,
\end{equation}
subject to normalization in $x$ for each fixed $y$.

The variables $y$ are not intended to represent underlying physical states of the system. Rather, they serve as auxiliary parameters labeling hypothetical configurations in an inference problem. In this sense, $P(x \mid y)$ is a likelihood: it quantifies the response of the detector labeled by $x$ under the assumption that the configuration is $y$. The present framework is therefore not a hidden-variable theory, since no ontological status is assigned to $y$ and physical predictions are obtained only after marginalizing over these parameters.

In this formulation, $x$ labels detector outcomes, while $y$ parametrizes a space of hypotheses about the processes that give rise to those outcomes. The kernel $K(x,y)$ encodes the response of the detector system to different hypothetical configurations and is generically nonlocal in parameter space: a given configuration $y$ may contribute to multiple detector outcomes $x$.

The wavefunction acquires a direct operational interpretation by defining
\begin{equation}
\psi_x(y) := K(x,y),
\end{equation}
which represents the amplitude that the detector at $x$ clicks given a hypothetical configuration $y$. Localization is therefore a property of detector labeling rather than of an underlying trajectory.

The use of complex amplitudes, rather than real probabilities, follows from the requirement that alternative hypothetical configurations combine consistently under coarse graining. This consistency requires linear additivity at the amplitude level, leading naturally to interference and therefore to complex-valued kernels.

Observable probabilities arise only after assigning weights to configurations or histories. In the dynamical setting, this weighting is provided by the path-integral structure,
\begin{equation}
K(x)=\int \mathcal D y\; e^{\frac{i}{\hbar}S[y]}\,W_x[y],
\label{eq:path_integral_kernel_new}
\end{equation}
where $S[y]$ determines the relative weight of different histories and $W_x[y]$ encodes their compatibility with the detector outcome $x$. In standard formulations, $W_x[y]$ plays a role analogous to a boundary state or prior weighting over configurations, but here it is interpreted operationally as part of the detector response structure rather than as an independent wavefunction.

At this stage, $S[y]$ should not be interpreted as the action of a system evolving on a pre-existing spacetime geometry. Rather, it is a functional assigning relative weights to hypothetical histories in the inference problem. The notion of spacetime geometry, and in particular the metric structure entering gravitational dynamics, will emerge only at a later stage from the distinguishability properties of detector states, avoiding any circular dependence.

Histories for which the functional $S[y]$ is stationary interfere constructively and dominate the integral, so that the classical principle of least action appears as a concentration of inferential weight rather than as an ontologically primitive law.

The path-integral representation should be understood as a continuum limit of a more general composition rule for detector-response amplitudes. If the contribution of successive segments of a configuration is required to combine locally and consistently under refinement, then the weighting functional must be additive,
\begin{equation}
S[y_1 \circ y_2] = S[y_1] + S[y_2],
\end{equation}
which leads naturally to an exponential form for the amplitude. The use of complex phases ensures that alternative configurations combine linearly while preserving interference under coarse graining.

On the other hand, the Gaussian reference family chosen as a starting point for the detectors is selected by a maximum-entropy principle. Among all detector states compatible with fixed local mean and covariance, the Gaussian distribution maximizes entropy and therefore provides the least biased choice. It defines a natural vacuum detector class relative to which deformations may be described.

Under Wick rotation, one obtains
\begin{equation}
P[y]\propto e^{-\frac{1}{\hbar}S_E[y]},
\end{equation}
which can be interpreted in the spirit of maximum-entropy methods as a least-biased assignment over histories subject to constraints encoded in the Euclidean action \cite{Jaynes1957,Caticha2012}. In the present work, however, the path-integral representation will play mainly an interpretive role. The geometric construction is developed directly from the detector family itself.

To proceed, one must introduce a reference family of detector states corresponding to the absence of external influences. We define the vacuum detector family as the least biased set of detector responses consistent with a minimal set of local constraints. Following the principle of maximum entropy, this leads naturally to Gaussian detector states. The point is not that Gaussians are fundamental, but that they are the least biased local detector profiles compatible with fixed first and second moments.

For the spatial detector sector, we therefore consider the normalized family of pure states
\begin{equation}
\psi_{\mathbf q}(\mathbf y)
=
\mathcal N(A)\,
\exp\!\left[
-\frac{1}{4}(\mathbf y-\mathbf q)^T A(\mathbf y-\mathbf q)
+\frac{i}{2}(\mathbf y-\mathbf q)^T B(\mathbf y-\mathbf q)
+i\,\mathbf p\cdot(\mathbf y-\mathbf q)
+i\phi
\right],
\label{eq:full_gaussian_detector_new}
\end{equation}
where $\mathbf q\in\mathbb R^3$ is the detector label, $\mathbf y\in\mathbb R^3$ is the auxiliary parameter variable, $A=A^T>0$ controls localization width, $B=B^T$ is a real symmetric phase-curvature matrix, $\mathbf p$ is a linear phase gradient, and $\phi$ is an overall phase. The normalization factor is
\begin{equation}
\mathcal N(A)=\left[(2\pi)^3\det(A^{-1})\right]^{-1/4}.
\end{equation}
The corresponding probability density is
\begin{equation}
|\psi_{\mathbf q}(\mathbf y)|^2
=
\frac{\sqrt{\det A}}{(2\pi)^{3/2}}
\exp\!\left[
-\frac{1}{2}(\mathbf y-\mathbf q)^T A(\mathbf y-\mathbf q)
\right].
\end{equation}
Thus the covariance matrix of the probability density is $A^{-1}$, while the matrix $B$ leaves the density unchanged but modifies the phase structure of the detector state.

For the temporal sector, we introduce the analogous one-dimensional family
\begin{equation}
\chi_t(\tau)
=
\mathcal N(a)\,
\exp\!\left[
-\frac{a}{4}(\tau-t)^2
+\frac{i b}{2}(\tau-t)^2
+i\pi(\tau-t)
+i\varphi
\right],
\label{eq:temporal_detector_new}
\end{equation}
with $a>0$ and $b,\pi,\varphi\in\mathbb R$.

The family \eqref{eq:full_gaussian_detector_new}--\eqref{eq:temporal_detector_new} will serve as the reference Gaussian detector class. The case $B=0$, $\mathbf p=0$ reduces to the simplest localized Gaussian detector. The more general phase structure is needed to discuss squeezing, phase holonomy, and the relation between detector modification and geometry.

The detector family defines a quantum statistical manifold. Its local geometry is encoded in the quantum geometric tensor. For a normalized family of pure states $|\psi(\xi)\rangle$, this is
\begin{equation}
Q_{AB}
=
\langle \partial_A \psi \mid \partial_B \psi \rangle
-
\langle \partial_A \psi \mid \psi \rangle
\langle \psi \mid \partial_B \psi \rangle.
\label{eq:QGT_new}
\end{equation}

Its real part defines the Fubini--Study metric, while its imaginary part defines the Berry curvature associated with phase holonomy \cite{Berry1984,BraunsteinCaves1994,Petz1996}. These correspond respectively to the Riemannian and symplectic structures naturally defined on the projective Hilbert space of pure states.

We begin with translations of the spatial detector label $\mathbf q$. For the detector state \eqref{eq:full_gaussian_detector_new},
\begin{equation}
\partial_{q_i}\psi_{\mathbf q}
=
\left[
\frac{1}{2}\bigl(A(\mathbf y-\mathbf q)\bigr)_i
-i\bigl(B(\mathbf y-\mathbf q)\bigr)_i
-i p_i
\right]\psi_{\mathbf q}.
\end{equation}
Using the Gaussian averages
\begin{equation}
\langle y_i-q_i\rangle = 0,
\qquad
\langle (y_i-q_i)(y_j-q_j)\rangle = (A^{-1})_{ij},
\end{equation}
one finds
\begin{equation}
\langle \psi_{\mathbf q}\mid \partial_{q_i}\psi_{\mathbf q}\rangle = -i p_i
\end{equation}
and
\begin{equation}
\langle \partial_{q_i}\psi_{\mathbf q}\mid \partial_{q_j}\psi_{\mathbf q}\rangle
=
\frac{1}{4}A_{ij}
+
(BA^{-1}B)_{ij}
+
p_i p_j.
\end{equation}
Substituting into \eqref{eq:QGT_new} gives the spatial distinguishability metric
\begin{equation}
g^{(\mathrm S)}_{ij}
=
4\,\mathrm{Re}\,Q_{ij}
=
A_{ij}+4(BA^{-1}B)_{ij}.
\label{eq:spatial_metric_general_new}
\end{equation}

This is a central result. It shows explicitly that detector-induced geometry depends not only on localization width, through $A$, but also on quadratic phase structure, through $B$. The phase structure of detector states therefore contributes to geometry even when it leaves the probability density unchanged.

The Berry connection on detector-state space is
\begin{equation}
\mathcal A_A = i\langle \psi \mid \partial_A \psi\rangle,
\end{equation}
and the Berry curvature is
\begin{equation}
\mathcal F_{AB}
=
\partial_A\mathcal A_B - \partial_B\mathcal A_A
=
2\,\mathrm{Im}\,Q_{AB}.
\end{equation}
Thus the detector manifold carries two coupled geometric structures: the distinguishability metric via
$\mathrm{Re}\,Q_{AB}$ and the Berry curvature via $\mathrm{Im}\,Q_{AB}$.

For the temporal detector \eqref{eq:temporal_detector_new}, the same calculation gives
\begin{equation}
g^{(\mathrm T)} = a + 4\frac{b^2}{a}.
\label{eq:temporal_metric_general_new}
\end{equation}
The temporal distinguishability scale therefore depends on both the temporal width and temporal phase curvature.

\section{Reconstruction of the Lorentzian Metric}

The quantum informational metric tensor defines a Riemannian metric on detector-state space. The physical spacetime metric is not identified with this object directly, but reconstructed from it.

We assume that detector outcomes admit a sufficiently regular coarse-grained structure for the space of labels to be modeled as a smooth manifold. In such regimes, the space of detector labels acquires the structure of a differentiable manifold. A spacetime point is then identified with a detector outcome label, together with the associated detector state. 

We distinguish spatial and temporal detector sectors and write
\begin{equation}
\xi^A = (\xi^a_{\mathrm S}, \xi^\alpha_{\mathrm T}).
\end{equation}
The spatial metric is defined by the pullback of the spatial block of the detector-state metric,
\begin{equation}
h_{ij}(x)
=
\mathcal G^{(\mathrm S)}_{ab}(\xi(x))\,\partial_i \xi^a_{\mathrm S}\,\partial_j \xi^b_{\mathrm S},
\end{equation}
while the temporal distinguishability scale is
\begin{equation}
N^2(x)
=
\mathcal G^{(\mathrm T)}_{\alpha\beta}(\xi(x))\,\partial_0 \xi^\alpha_{\mathrm T}\,\partial_0 \xi^\beta_{\mathrm T}.
\end{equation}

The Lorentzian metric is then reconstructed as
\begin{equation}
ds^2 = -N^2(x)\,dt^2 + h_{ij}(x)\,dx^i dx^j.
\label{eq:lorentz_reconstruct}
\end{equation}

This construction should be understood as a reconstruction ansatz valid in detector-adapted frames. The detector-state geometry itself is positive-definite, but the spatial and temporal detector sectors play qualitatively different operational roles. The spatial sector is associated with localization within a simultaneity slice, whereas the temporal sector is associated with clock readings and temporal ordering. In the detector-adapted regime considered here, the temporal detector family defines a preferred local congruence orthogonal to the spatial detector slices, so that the mixed components $g_{0i}$ vanish and the metric takes the block-diagonal form \eqref{eq:lorentz_reconstruct}. More general configurations, including nonzero shift terms, arise when this orthogonality is relaxed or when one passes to non-adapted frames, but they are not required for the present analysis.

The Lorentzian signature is therefore not inherited directly from the detector-state metric, but introduced at the reconstruction stage through the distinct operational interpretation of the temporal detector sector. In this sense, the Euclidean formulation may be viewed as an analytic continuation of the temporal sector, analogous to the usual Wick rotation, but not as a consequence of any literal orthogonality between real and imaginary numbers.

At the effective level, this means that Wick rotation acts on the reconstructed temporal ordering variable rather than on a fundamental spacetime background, since the latter has not yet been assumed at the detector level.

\section{Connection, Curvature, and Calibration Transport}

Once the metric is reconstructed, one obtains the Levi--Civita connection
\begin{equation}
\Gamma^\lambda{}_{\mu\nu}
=
\frac{1}{2} g^{\lambda\rho}
\left(
\partial_\mu g_{\nu\rho}
+
\partial_\nu g_{\mu\rho}
-
\partial_\rho g_{\mu\nu}
\right).
\end{equation}

This connection has a direct operational meaning. It specifies how local detector calibrations are compared across spacetime while preserving local distinguishability. Parallel transport corresponds to comparing detector calibrations between neighboring events in such a way that their local response properties are preserved. This should not be interpreted as physically moving a detector through spacetime. Rather, it is a rule for identifying or relabeling detector responses at nearby points while maintaining equivalence of their local distinguishability structure.

In this sense, parallel transport defines a consistency condition for how detector calibrations are matched across the event manifold. When torsion is absent, this identification is path-independent to first order, and the notion of transporting a calibration between neighboring events is unambiguous. The Levi--Civita connection therefore encodes not motion, but the compatibility of local detector-response structures under changes of label.

The failure of path independence is measured by curvature. For a small loop,
\begin{equation}
\delta V^\alpha
=
R^\alpha{}_{\beta\mu\nu} V^\beta \delta S^{\mu\nu}.
\end{equation}
Thus the Riemann tensor measures calibration holonomy.

The operational interpretation of curvature developed above can be sharpened into a justification of the geometric sector of the effective action. The question is not merely which tensor measures infinitesimal calibration holonomy, but which local scalar density measures the extent to which neighboring detector calibrations fail to patch into a globally consistent distinguishability structure.

To identify that scalar, consider a geodesic ball $B_r(x)$ of radius $r$ centered at an event $x$, defined with respect to the reconstructed metric. In Riemann normal coordinates about $x$, the metric has the expansion
\begin{equation}
g_{\mu\nu}(y)
=
\eta_{\mu\nu}
-
\frac{1}{3}R_{\mu\alpha\nu\beta}(x)\,y^\alpha y^\beta
+
\mathcal O(y^3),
\label{eq:RNC_metric_paper}
\end{equation}
and therefore the invariant volume density expands as
\begin{equation}
\sqrt{|g(y)|}
=
1
-
\frac{1}{6}R_{\alpha\beta}(x)\,y^\alpha y^\beta
+
\mathcal O(y^3).
\label{eq:sqrtg_RNC_paper}
\end{equation}
Integrating over the ball gives the standard small-radius expansion
\begin{equation}
\mathrm{Vol}(B_r(x))
=
\omega_n r^n
\left[
1
-
\frac{R(x)}{6(n+2)}\,r^2
+
\mathcal O(r^3)
\right],
\label{eq:ball_volume_expansion_paper}
\end{equation}
where $n$ is the dimension and $\omega_n$ is the Euclidean unit-ball volume.

Within the present framework, this expansion has a direct operational meaning. The invariant measure $\sqrt{|g|}\,d^n x$ counts the local density of distinguishable detector outcomes. Accordingly, $\mathrm{Vol}(B_r(x))$ measures the number of outcomes accessible within operational distance $r$ from the event $x$. Curvature changes this number relative to the flat case. The leading deviation is controlled by the scalar curvature $R(x)$.

This identifies $R(x)$ as the leading local scalar measure of detector-patching inconsistency. Equivalently, scalar curvature is the unique lowest-order trace measure of how local calibration holonomy modifies the density of distinguishable events.

This point may also be expressed from the perspective of geodesic congruences. The Riemann tensor measures loop holonomy, but the Ricci tensor and its scalar trace arise when one asks how calibration mismatch affects the convergence or divergence of neighboring geodesics and therefore the packing of distinguishable detector outcomes. In this sense, the scalar curvature is not merely an algebraic contraction of the Riemann tensor. It is the local scalar quantifying how calibration inconsistency changes the effective density of operationally distinguishable events.

Once this interpretation is established, the form of the leading geometric contribution to the effective functional is strongly constrained. The action must be an integral over spacetime of a local scalar density measuring detector-patching inconsistency. The invariant measure is $\sqrt{|g|}\,d^4x$, and the leading local scalar built from the metric and two derivatives is $R[g]$. Hence the geometric sector must take the form
\begin{equation}
I_{\mathrm{geom}}[g]
=
\alpha \int d^4x\,\sqrt{|g|}\,R[g]
\label{eq:geom_functional_final}
\end{equation}

up to the addition of a cosmological term proportional to $\int d^4x\,\sqrt{|g|}$, which we set to zero for simplicity. In the present framework, such a term would correspond to a uniform offset in the density of distinguishable detector outcomes, i.e. a baseline calibration of the detector network that does not affect local distinguishability relations. Since it contributes only a constant background to the effective geometry, it can be consistently neglected at this stage.

The logic here is effective rather than microscopic. The detector framework does not yet derive the Ricci scalar directly from the kernel without passing through smooth geometry. What it does show is that, once detector distinguishability defines a smooth Lorentzian metric and curvature is interpreted as detector-patching inconsistency, the scalar curvature is the unique leading local scalar encoding that inconsistency.

This uniqueness may be sharpened in two ways. First, diffeomorphism invariance requires the geometric contribution to be a scalar density. Second, if one restricts attention to local functionals involving at most two derivatives of the metric and demands second-order field equations, then the Einstein--Hilbert term is uniquely selected, up to said cosmological constant term and higher-derivative corrections. In four dimensions this is the standard Lovelock uniqueness statement specialized to the leading low-energy sector.

Accordingly, the Einstein--Hilbert term is not inserted here by mere analogy with general relativity. It is selected as the universal leading consistency functional of the reconstructed detector geometry. The metric is reconstructed from detector distinguishability. The Riemann tensor measures infinitesimal calibration holonomy. The Ricci scalar measures the leading local deficit in the density of distinguishable outcomes caused by that holonomy. Integrating $\sqrt{|g|}R$ over spacetime accumulates this inconsistency density globally. This is the precise sense in which the geometric sector of the effective action emerges from detector patching.

\section{Detector Deformation and Induced Geometry}

Matter enters the framework as local deformation of the detector family away from the vacuum Gaussian class. In the following, variations are taken with respect to the inverse metric $g^{\mu\nu}$. The variation of the determinant is therefore
\[
\delta \sqrt{|g|}
=
-\frac{1}{2}\sqrt{|g|}\,g_{\mu\nu}\,\delta g^{\mu\nu}.
\]

The most general small deformation of the detector model \eqref{eq:full_gaussian_detector_new}--\eqref{eq:temporal_detector_new} changes the Gaussian parameters:
\begin{equation}
A \rightarrow A+\delta A,
\qquad
B \rightarrow B+\delta B,
\qquad
\mathbf p \rightarrow \mathbf p+\delta \mathbf p,
\qquad
a \rightarrow a+\delta a,
\qquad
b \rightarrow b+\delta b.
\label{eq:detector_deformation}
\end{equation}

Since the spacetime metric is reconstructed from the detector family, any deformation of the detector parameters induces a corresponding metric variation. If $\xi^A(x)$ denotes the full set of detector-state parameters, then
\begin{equation}
\delta g_{\mu\nu}
=
\frac{\partial g_{\mu\nu}}{\partial \xi^A}\,\delta \xi^A.
\label{eq:full_metric_variation}
\end{equation}
This expression makes explicit that spatial and temporal sectors are not independent. A deformation of the detector family generically affects both spatial and temporal components of the reconstructed metric.

For the spatial sector,
\begin{equation}
g^{(\mathrm S)} = A + 4BA^{-1}B,
\end{equation}
so that
\begin{equation}
\delta g^{(\mathrm S)}
=
\delta A
+
4\left(
\delta B A^{-1}B
+
B A^{-1}\delta B
-
B A^{-1}\delta A A^{-1}B
\right).
\label{eq:delta_spatial_metric_full}
\end{equation}

For the temporal sector,
\begin{equation}
g^{(\mathrm T)} = a + 4\frac{b^2}{a},
\end{equation}
giving
\begin{equation}
\delta g^{(\mathrm T)}
=
\delta a
+
8\frac{b}{a}\delta b
-
4\frac{b^2}{a^2}\delta a.
\label{eq:delta_temporal_metric_full}
\end{equation}

Equations \eqref{eq:delta_spatial_metric_full} and \eqref{eq:delta_temporal_metric_full} show that geometry responds to both amplitude deformations and phase deformations. In particular, phase curvature alone can generate nontrivial metric structure even when the probability distribution remains unchanged.

This point can be illustrated explicitly. Consider the case
\begin{equation}
A_0=\sigma_0^{-2}\delta_{ij},
\qquad
B_{ij}(x)=\beta(x)\delta_{ij}.
\end{equation}
Then
\begin{equation}
g^{(\mathrm S)}_{ij}
=
\left(
\sigma_0^{-2}+4\sigma_0^2\beta(x)^2
\right)\delta_{ij}.
\end{equation}
Although the probability distribution is spatially uniform, the metric becomes position dependent whenever $\beta(x)$ varies. Curvature therefore arises entirely from phase structure. This sharply distinguishes the present construction from purely classical information-geometric approaches in which geometry depends only on probability distributions.

Thus matter, understood as detector deformation, is not merely a broadening or narrowing of detector widths. It also includes changes in phase structure, squeezing, and coherence properties. The induced metric responds to the full quantum structure of the detector family.

\section{Consistency Functional and Coupled Field Equations}

The geometric sector of the theory has been identified in the previous section. There, scalar curvature was shown to measure the local deficit of distinguishable detector outcomes arising from calibration holonomy. The corresponding global consistency cost is therefore given by the Einstein--Hilbert term.

To describe the coupled dynamics of geometry and detector deformation, we introduce an effective consistency functional built from three ingredients.

First, local deformation of the detector family away from the vacuum reference state $\rho^{(0)}$ carries a cost measured by the quantum relative entropy
\begin{equation}
S\!\left(\rho(\xi)\middle\|\rho^{(0)}\right),
\end{equation}
where $\rho(\xi)$ denotes the detector state parametrized by deformation variables $\xi^A$. We define the corresponding local potential
\begin{equation}
U(\xi):=\mu\,S\!\left(\rho(\xi)\middle\|\rho^{(0)}\right),
\end{equation}
with $\mu$ a phenomenological coupling constant setting the scale of the deformation cost.

Second, variation of detector states across neighboring events is measured by the pullback of the detector-state metric $\mathcal G_{AB}(\xi)$, defined as the real part of the quantum geometric tensor introduced above,
\begin{equation}
\mathcal G_{AB}(\xi) := 4\,\mathrm{Re}\,Q_{AB}(\xi),
\end{equation}
so that $\mathcal G_{AB}$ provides the distinguishability metric on detector-state space. In particular, the spatial and temporal metrics constructed in the previous section correspond to blocks of $\mathcal G_{AB}$ associated with translations in the respective detector sectors.

This gives the nearest-neighbor consistency term
\begin{equation}
\frac{\nu}{2}\,\mathcal G_{AB}(\xi)\,g^{\mu\nu}\partial_\mu \xi^A \partial_\nu \xi^B,
\end{equation}
where $\nu$ is a second phenomenological coupling constant.

Third, the geometric sector is described by the Einstein--Hilbert term, selected at the effective level as the unique leading local scalar functional of the reconstructed metric compatible with locality, diffeomorphism invariance, and second-order dynamics.

We therefore consider
\begin{equation}
\mathcal I[\xi,g]
=
\int d^4x\,\sqrt{|g|}
\left[
\alpha R[g]
+
U(\xi)
+
\frac{\nu}{2}\,\mathcal G_{AB}(\xi)\,g^{\mu\nu}\partial_\mu \xi^A \partial_\nu \xi^B
\right],
\label{eq:masterfunctional_clean}
\end{equation}
where $\alpha$ is the gravitational coupling. The first term measures geometric inconsistency of detector patching, the second the local cost of deforming detector states away from the vacuum family, and the third the cost of varying those deformations across spacetime.

We therefore consider
\begin{equation}
\mathcal I[\xi,g]
=
\int d^4x\,\sqrt{|g|}
\left[
\alpha R[g]
+
U(\xi)
+
\frac{\nu}{2}\,\mathcal G_{AB}(\xi)\,g^{\mu\nu}\partial_\mu \xi^A \partial_\nu \xi^B
\right],
\label{eq:masterfunctional_clean}
\end{equation}

and derive the coupled field equations from it.

Variation with respect to $\xi^A$ gives the deformation-field equation. Since the Einstein--Hilbert term does not depend explicitly on $\xi^A$, one varies the deformation functional
\begin{equation}
I_{\mathrm{def}}[\xi,g]
=
\int d^4x\,\sqrt{|g|}
\left[
U(\xi)
+
\frac{\nu}{2}\,\mathcal G_{AB}(\xi)\,g^{\mu\nu}\partial_\mu \xi^A \partial_\nu \xi^B
\right].
\end{equation}
The corresponding Lagrangian density is
\begin{equation}
\mathcal L
=
\sqrt{|g|}
\left[
U(\xi)
+
\frac{\nu}{2}\,\mathcal G_{AB}(\xi)\,g^{\mu\nu}\partial_\mu \xi^A \partial_\nu \xi^B
\right].
\end{equation}
One computes
\begin{equation}
\frac{\partial \mathcal L}{\partial(\partial_\mu \xi^C)}
=
\sqrt{|g|}\,\nu\,\mathcal G_{CB}(\xi)\,g^{\mu\nu}\partial_\nu \xi^B
\end{equation}
and
\begin{equation}
\frac{\partial \mathcal L}{\partial \xi^C}
=
\sqrt{|g|}
\left[
\partial_C U(\xi)
+
\frac{\nu}{2}(\partial_C \mathcal G_{AB})\,g^{\mu\nu}\partial_\mu \xi^A\partial_\nu \xi^B
\right].
\end{equation}
Substituting into the Euler--Lagrange equation and rewriting the result in covariant form yields

\begin{equation}
\nu\left[
\Box_g \xi^A
+
\Gamma^A{}_{BC}(\mathcal G)\,g^{\mu\nu}\partial_\mu \xi^B \partial_\nu \xi^C
\right]
-
\mathcal G^{AB}\partial_B U
=
0.
\label{eq:sigma_eq}
\end{equation}
where $\Gamma^A{}_{BC}(\mathcal G)$ is the Levi--Civita connection of detector-state space.

Equation \eqref{eq:sigma_eq} describes the evolution of detector deformations as a nonlinear sigma model on the manifold of detector states. The first term describes propagation on reconstructed spacetime, the second reflects the internal curvature of detector-state space, and the third is the restoring force associated with the local deformation cost.

The reconstruction outlined above relies on a set of structural assumptions that define the domain of validity of the framework. First, we assume that detector outcomes admit a smooth coarse-grained parametrization, so that the space of detector labels can be treated as a differentiable manifold. Second, we restrict attention to detector-adapted frames in which the metric takes the block-diagonal form \eqref{eq:lorentz_reconstruct}, so that shift terms are absent at leading order. Third, the Lorentzian signature is not derived from the detector-state geometry, which is positive-definite, but arises from the identification of one detector sector with temporal ordering.

These assumptions should be understood as defining an effective regime in which detector families are sufficiently regular to admit a geometric description. The resulting theory is therefore an effective description of detector response structure at scales where such a smooth reconstruction is valid.

We now vary the action with respect to the metric while holding $\xi^A$ fixed. The variation of the Einstein--Hilbert term is
\begin{equation}
\delta\!\left(\sqrt{|g|}R\right)
=
\sqrt{|g|}\,G_{\mu\nu}\,\delta g^{\mu\nu}
+
\sqrt{|g|}\,\nabla_\lambda V^\lambda,
\end{equation}
where the final term is a boundary contribution. Restricting attention to bulk variations,
\begin{equation}
\delta I_{\mathrm{grav}}
=
\alpha \int d^4x\,\sqrt{|g|}\,G_{\mu\nu}\,\delta g^{\mu\nu}.
\end{equation}

For the deformation sector, one uses
\begin{equation}
\delta\sqrt{|g|}=-\frac12 \sqrt{|g|}\,g_{\mu\nu}\,\delta g^{\mu\nu}
\end{equation}
to obtain
\begin{equation}
\delta I_{\mathrm{def}}
=
-\frac12 \int d^4x\,\sqrt{|g|}\,T_{\mu\nu}^{(\xi)}\,\delta g^{\mu\nu},
\end{equation}
with
\begin{equation}
T_{\mu\nu}^{(\xi)}
=
\nu\,\mathcal G_{AB}(\xi)\,\partial_\mu \xi^A \partial_\nu \xi^B
-
g_{\mu\nu}
\left[
U(\xi)
+
\frac{\nu}{2}\,\mathcal G_{AB}(\xi)\,g^{\rho\sigma}\partial_\rho \xi^A \partial_\sigma \xi^B
\right].
\label{eq:Tmunu_xi}
\end{equation}
This tensor is defined by variation of the deformation sector of the functional with respect to the metric,
\begin{equation}
T_{\mu\nu}^{(\xi)}
=
-\frac{2}{\sqrt{|g|}}\frac{\delta I_{\mathrm{def}}}{\delta g^{\mu\nu}},
\end{equation}

and therefore represents the response of the detector-deformation cost to local changes in the reconstructed geometry. In this sense it plays the role of a stress-energy tensor.

Operationally, $T_{\mu\nu}^{(\xi)}$ quantifies how the cost of maintaining detector consistency is distributed across spacetime: the first term measures directional variation of detector deformations, while the second subtracts the local scalar density of deformation cost. Its appearance as a source in the metric equation expresses the fact that nonuniform detector deformations modify the reconstructed geometry.

Stationarity of the full functional then gives
\begin{equation}
\alpha G_{\mu\nu}
=
\frac12 T_{\mu\nu}^{(\xi)}.
\end{equation}
Fixing
\begin{equation}
\alpha=\frac{1}{16\pi G}
\end{equation}
yields the Einstein equation
\begin{equation}
G_{\mu\nu}=8\pi G\,T_{\mu\nu}^{(\xi)}.
\label{eq:einstein_final}
\end{equation}

The meaning of \eqref{eq:einstein_final} within the present framework deserves emphasis. The metric \(g_{\mu\nu}\) is not introduced as a fundamental background variable, but reconstructed from the distinguishability structure of the detector family. The Einstein equation is therefore not interpreted here as a microscopic law imposed on spacetime itself. Rather, it is the stationarity condition of the coarse-grained consistency functional \eqref{eq:masterfunctional_clean}. It states that the global geometry reconstructed from detector responses is admissible only when its Einstein tensor is balanced by the stress-energy associated with detector deformation.

At this stage, the effective geometric description has been formulated independently of the underlying weighting functional $S[y]$. The relation between these two levels will be clarified in the following section.

This derivation also sharpens the distinction between vacuum and non-vacuum sectors. In the vacuum sector,
\begin{equation}
\xi^A=\mathrm{const.},
\qquad
\partial_\mu \xi^A=0,
\qquad
U(\xi)=0,
\end{equation}
so that
\begin{equation}
T_{\mu\nu}^{(\xi)}=0,
\end{equation}
and the field equation reduces to
\begin{equation}
G_{\mu\nu}=0.
\end{equation}
Thus vacuum solutions need not be flat. They are precisely those geometries that arise from undeformed local detector families whose global patching remains nontrivial.

Finally, since the total action is diffeomorphism invariant, the covariant conservation law
\begin{equation}
\nabla^\mu T_{\mu\nu}^{(\xi)}=0
\end{equation}
follows from the contracted Bianchi identity. This expresses the consistency of detector deformation dynamics with the reconstructed spacetime geometry.

\section{Effective Matter Sector}

The effective geometric description introduced above may be viewed as a coarse-grained representation of the underlying weighting structure defined by the functional $S[y]$. At the kinematical level, $S[y]$ assigns relative weights to hypothetical histories without reference to any pre-existing spacetime geometry.

In regimes where detector families admit a smooth geometric representation, consistency requirements on detector responses select an effective spacetime metric and associated curvature. The resulting dynamics is then captured by a local action whose leading term is the Einstein--Hilbert functional, while the remaining terms encode detector deformations.

The variables $\xi^A(x)$ in the master functional are coordinates on detector-state deformation space. They are not fundamental matter fields. To recover familiar field-theoretic structure, one coarse grains around the vacuum detector family.

Let $\rho^{(0)}=\rho(0)$ denote the reference state and expand for small deformations:
\begin{equation}
\rho(\xi)=\rho^{(0)}+\xi^A \partial_A \rho\big|_0 + \frac{1}{2}\xi^A\xi^B \partial_A\partial_B \rho\big|_0 + \cdots.
\end{equation}
The local relative-entropy cost admits the quadratic expansion
\begin{equation}
S\!\left(\rho(\xi)\middle\|\rho^{(0)}\right)
=
\frac{1}{2}M_{AB}\,\xi^A\xi^B + \mathcal O(\xi^3),
\label{eq:relative_entropy_expand}
\end{equation}
where $M_{AB}$ is the Hessian at the reference family. Likewise,
\begin{equation}
\mathcal G_{AB}(\xi)=\mathcal G^{(0)}_{AB}+\mathcal O(\xi).
\end{equation}
To quadratic order, the effective action becomes
\begin{equation}
\mathcal I_{\mathrm{eff}}[\xi,g]
=
\int d^4x\,\sqrt{|g|}
\left[
\alpha R[g]
+
\frac{1}{2}\mu\,M_{AB}\,\xi^A\xi^B
+
\frac{\nu}{2}\,\mathcal G^{(0)}_{AB}\,g^{\mu\nu}\partial_\mu \xi^A \partial_\nu \xi^B
\right].
\label{eq:quadratic_xi_action}
\end{equation}
This is the action of a multiplet of scalar fields on curved spacetime, with mass matrix \(M_{AB}\) and kinetic matrix \(\mathcal G^{(0)}_{AB}\).

The simplest truncation is the one-dimensional deformation sector
\begin{equation}
\xi^A(x)\to \phi(x),
\end{equation}
representing, for instance, an isotropic deformation of the detector family. Then
\begin{equation}
\mathcal G_{AB}\to \chi(\phi),
\end{equation}
and the master functional reduces to
\begin{equation}
\mathcal I[\phi,g]
=
\int d^4x\,\sqrt{|g|}
\left[
\alpha R[g]
+
U(\phi)
+
\frac{\nu}{2}\chi(\phi)\,g^{\mu\nu}\partial_\mu\phi\,\partial_\nu\phi
\right].
\label{eq:scalar_master}
\end{equation}
Near the vacuum,
\begin{equation}
U(\phi)=\frac{1}{2}m^2\phi^2+\lambda_3\phi^3+\lambda_4\phi^4+\cdots,
\qquad
\chi(\phi)=\chi_0+\chi_1\phi+\cdots.
\end{equation}
After the field redefinition
\begin{equation}
\varphi = \sqrt{\nu\chi_0}\,\phi,
\end{equation}
the leading-order action becomes
\begin{equation}
\mathcal I[\varphi,g]
=
\int d^4x\,\sqrt{|g|}
\left[
\alpha R[g]
+
\frac{1}{2}g^{\mu\nu}\partial_\mu \varphi \partial_\nu \varphi
+
V(\varphi)
\right].
\label{eq:canon_scalar}
\end{equation}
Thus a standard scalar field theory emerges as the low-energy effective description of a one-parameter detector-deformation sector.

Within the present framework, the scalar field \(\varphi\) is not a fundamental substance propagating on spacetime. It is the coarse-grained variable describing how the detector family departs from its vacuum reference configuration. Its stress-energy is the operational cost of detector deformation distributed across spacetime.

The same logic extends, in principle, to higher-dimensional deformation sectors. Different effective matter multiplets correspond to different local deformation modes of the detector family. The present paper does not attempt a full classification of such sectors; it establishes only the general mechanism by which field-theoretic dynamics emerges.

In this way, the pre-geometric weighting functional $S[y]$ provides the underlying inferential structure, while the effective spacetime action emerges as its coarse-grained geometric representation.

\section{Discussion}

The framework developed in this work proposes a unified perspective in which quantum theory and spacetime geometry arise from a common inferential structure defined on the space of detector outcomes. The central elements of this construction are the detector kernel, the detector family, and the geometry of distinguishability on detector-state space. Within this setting, spacetime geometry is not introduced as a primitive structure, but is reconstructed from the organization of detector responses, while effective matter fields correspond to deformations of this response structure.

A distinctive feature of the framework is the explicit separation between detector-state geometry and spacetime geometry. The detector-state manifold carries the quantum geometric tensor, whose real part gives a metric of distinguishability and whose imaginary part gives a Berry connection and its associated phase holonomy. The reconstructed spacetime carries a different connection, namely the Levi--Civita connection of the Lorentzian metric, whose holonomy measures the failure of local detector calibrations to be globally consistent. This double geometric structure is one of the main conceptual tools of the analysis.

Equally important is the distinction between vacuum geometry and matter. Vacuum does not mean flatness, but rather that the detector family remains within the reference Gaussian class, without local deformation fields. Curvature may still be nonzero because the global patching of locally admissible detector states may be nontrivial. Matter enters only when the detector family is locally deformed. In that case, the induced metric changes explicitly through the variation of Gaussian parameters, and the deformation fields source the effective Einstein equation.

The use of Gaussian detector families is motivated by the maximum-entropy principle and provides the simplest explicit setting in which the framework can be carried through analytically. The resulting formulas already show that geometry depends on both localization width and phase structure. In particular, phase-curvature parameters affect the induced metric and therefore must be treated as genuine contributors to geometry rather than as merely auxiliary quantum phases.

The present construction is effective in character. The master functional \eqref{eq:masterfunctional_clean} is motivated by distinguishability, locality, and diffeomorphism invariance, but it is not derived here from a microscopic theory of detectors. The Einstein--Hilbert term therefore has the same status as in effective field theory treatments of gravity: it is the leading local scalar term of the coarse-grained geometric sector \cite{Donoghue1994,Burgess2004}.

In this sense, the Einstein--Hilbert term is not postulated as a fundamental law, but identified as the universal leading contribution to the consistency cost of detector geometry. Potential observable deviations arise generically from higher-order corrections to the detector-state metric or from non-Gaussian detector sectors, leading to modifications of geodesic propagation or dispersion relations. For instance, phase-curvature-induced modifications of the detector metric generically lead to corrections of the effective dispersion relation for propagating excitations, providing a potential observational signature of the framework.

Another significant feature of the framework is that curvature may arise from phase structure alone. In standard information-geometric approaches, geometry is derived from probability distributions and therefore depends only on statistical distinguishability. In contrast, the present construction shows that quantum phase structure, encoded in the Gaussian phase-curvature parameters and, more generally, in the imaginary part of the quantum geometric tensor, affects the induced geometry through detector deformation. This indicates that spacetime geometry is sensitive not only to classical statistical structure, but also to genuinely quantum aspects of detector response.

Several limitations should be kept in view. The analysis assumes smooth detector families described by a finite number of parameters. It also assumes that detector sectors can be identified sufficiently well to reconstruct Lorentzian signature. In regimes where detector responses become highly nonlocal or strongly non-Gaussian, the effective metric description may cease to be adequate, and a more fundamental treatment in terms of the full detector kernel may be required.

Despite these limitations, the framework provides a coherent conceptual picture in which otherwise disparate elements of modern physics are brought into alignment. Quantum theory is understood as a theory of inference over detector outcomes, spacetime geometry as a measure of distinguishability together with calibration consistency, and effective matter fields as coarse-grained detector deformations. The present work is intended as a foundation for further development, and it suggests that a significant part of the formal content of both quantum theory and gravitation may be understood from the response structure of detectors.

\end{document}